\documentclass{emulateapj}

\usepackage{graphicx}
\usepackage{float}
\usepackage{amsmath}
\usepackage{epsfig,floatflt}

\begin{document}

\title{CMB likelihood approximation by a Gaussianized Blackwell-Rao estimator}

\author{\O. Rudjord\altaffilmark{1,4},
  N. E. Groeneboom\altaffilmark{2,4}, H. K. Eriksen\altaffilmark{3,4,5},
  Greg Huey\altaffilmark{6}, K. M. G\'{o}rski\altaffilmark{6,7} and J. B. Jewell\altaffilmark{6}}

\altaffiltext{1}{email: oystein.rudjord@astro.uio.no}
\altaffiltext{2}{email: leuat@irio.co.uk (NEG)}
\altaffiltext{3}{email: h.k.k.eriksen@astro.uio.no}

\altaffiltext{4}{Institute of Theoretical Astrophysics, University of
Oslo, P.O.\ Box 1029 Blindern, N-0315 Oslo, Norway}

\altaffiltext{5}{Centre of
Mathematics for Applications, University of Oslo, P.O.\ Box 1053
Blindern, N-0316 Oslo}

\altaffiltext{6}{Jet Propulsion Laboratory, California Institute
of Technology, Pasadena CA 91109} 

\altaffiltext{7}{Warsaw University Observatory, Aleje Ujazdowskie 4, 00-478 Warszawa,
  Poland}


\begin{abstract}
  We introduce a new CMB temperature likelihood approximation called
  the Gaussianized Blackwell-Rao (GBR) estimator. This estimator is
  derived by transforming the observed marginal power spectrum
  distributions obtained by the CMB Gibbs sampler into standard
  univariate Gaussians, and then approximate their joint transformed
  distribution by a multivariate Gaussian. The method is exact for
  full-sky coverage and uniform noise, and an excellent approximation
  for sky cuts and scanning patterns relevant for modern satellite
  experiments such as WMAP and Planck. The result is a stable,
  accurate and computationally very efficient CMB temperature
  likelihood representation that allows the user to exploit the unique
  error propagation capabilities of the Gibbs sampler to high
  $\ell$'s. A single evaluation of this estimator between $\ell =2$
  and 200 takes $\sim 0.2$ CPU milliseconds, while for comparison, a
  singe pixel space likelihood evaluation between $\ell = 2$ and 30
  for a map with $\sim2500$ pixels requires $\sim20$ seconds. We apply
  this tool to the 5-year WMAP temperature data, and re-estimate the
  angular temperature power spectrum, $C_{\ell}$, and likelihood,
  $\mathcal{L}(C_{\ell})$, for $\ell \le 200$, and derive new
  cosmological parameters for the standard six-parameter $\Lambda$CDM
  model. Our spectrum is in excellent agreement with the official WMAP
  spectrum, but we find slight differences in the derived cosmological
  parameters. Most importantly, the spectral index of scalar
  perturbations is $n_{\textrm{s}} = 0.973\pm 0.014$, $1.9\sigma$ away
  from unity and $0.6\sigma$ higher than the official WMAP result,
  $n_{\textrm{s}} = 0.965\pm 0.014$. This suggests that an exact
  likelihood treatment is required to higher $\ell$'s than previously
  believed, reinforcing and extending our conclusions from the 3-year
  WMAP analysis. In that case, we found that the sub-optimal
  likelihood approximation adopted between $\ell=12$ and 30 by the
  WMAP team biased $n_{\textrm{s}}$ low by $0.4\sigma$, while here we
  find that the same approximation between $\ell=30$ and 200
  introduces a bias of $0.6\sigma$ in $n_{\textrm{s}}$.
\end{abstract}


\section{Introduction}
\label{sec:introduction}

Detailed measurements of fluctuations in the cosmic microwave
background (CMB) have established cosmology as a high-precision
science. One striking illustration of this is the fact that it is
today possible to predict a vast number of observables based on six
numbers only, with only a few (but nevertheless intriguing)
``glitches'' overall. The key to this success has been making accurate
measurements of the CMB power spectrum, perhaps most prominently
exemplified by Wilkinson Microwave Anisotropy Probe (WMAP; Bennett et
al.\ 2003; Hinshaw et al.\ 2007, Hinshaw et al.\ 2008).

The primary connection between theoretical models and CMB observations
is made through the CMB likelihood, $\mathcal{L}(C_{\ell}) =
P(\mathbf{d}|C_{\ell})$. This is a multivariate, non-Gaussian function
that quantifies the match between the data and a given power spectrum,
$C_{\ell}$. Unfortunately, it is impossible to evaluate this function
explicitly for modern high-resolution data sets, due to the sheer size
of the problem, and one therefore instead typically resolves to
various approximations.

However, given the importance of the CMB in modern cosmology, it is of
critical importance to characterize this likelihood accurately, and
all approximations must be thoroughly verified. One example is the
approximation of the large angular scale likelihood, where
$\mathcal{L}(C_{\ell})$ is strongly non-Gaussian. This turned out to
be a non-trivial issue after the original analysis of the 3-year WMAP
temperature data by \citet{hinshaw:2007}, in which a Master-based
\citep{hivon:2002, verde:2003} approximation was used at $\ell > 12$.
An exact likelihood analysis \citep{eriksen:2007b} later demonstrated
that this sub-optimal approximation, when applied to harmonic modes
between $\ell=13$ and 30, biased the spectral index of scalar
perturbations, $n_{\textrm{s}}$, low by $0.4\sigma$.

A second example is that of non-cosmological foregrounds. Unless
properly accounted for, such foregrounds bias the observed power
spectrum to high values, and can seriously compromise any cosmological
conclusions. While important for temperature observations, this is an
absolutely crucial issue for polarization observations, as the desired
CMB in amplitude is comparable to or weaker than the interfering
foregrounds over most of the sky.

In recent years, a new set of statistical methods have been developed
that allows the user to address these issues within a single
well-defined framework \citep{jewell:2004, wandelt:2004,
  eriksen:2004}. The heart of this method is the Gibbs sampling
algorithm (see, e.g, Gelfand and Smith 1990), in which samples from a
(typically complicated) joint distribution are drawn by alternately
sampling from (simpler) conditional distributions. In the CMB setting,
this is realized by drawing joint samples from $P(\mathbf{s},
C_{\ell}|\mathbf{d})$, by alternately sampling from
$P(C_{\ell}|\mathbf{s}, \mathbf{d})$, where $C_{\ell}$ is the CMB
power spectrum, $\mathbf{s}$ is the CMB sky signal, and $\mathbf{d}$
are the observed data. In addition to allow for exact likelihood
analysis at reasonable computational cost, an equally important
feature of this framework is its unique capability of including
additional degrees of freedom, such as non-cosmological foregrounds,
into the analysis \citep{eriksen:2008a, eriksen:2008b}. Further, very
recently an additional Metropolis-Hastings MCMC sampling step was
introduced by \citet{jewell:2008}, that effectively resolves the
previously described inefficiency of the Gibbs sampler at low
signal-to-noise \citep{eriksen:2004}. The framework has also been
extended to handle polarization \citep{larson:2007, eriksen:2007b} and
anisotropic universe models \citep{groeneboom:2008}.

By now, the CMB Gibbs sampler is well established and demonstrated to
sample efficiently from the exact CMB posterior. However, a
long-standing issue has been the characterization of the joint
likelihood, given a set of such samples. Originally,
\citet{wandelt:2004} proposed to use the so-called Blackwell-Rao (BR)
estimator for this purpose, and this approach was later implemented
and studied in detail by \citet{chu:2005}. While highly accurate for
the large angular scale and high signal-to-noise temperature
likelihood, it suffers from one major drawback: Because it attempts to
describe the full $\ell_{\textrm{max}}$-dimensional likelihood without
any constraints on allowed correlations, the number of samples
required for convergence scales exponentially with
$\ell_{\textrm{max}}$. In practice, this limits the BR estimator to
$\ell \lesssim 30$ for temperature data, and just $\ell \lesssim 3-4$
for low signal-to-noise polarization data.

In this paper, we introduce a new temperature likelihood approximation
based on samples drawn from the CMB posterior, by modifying the
original BR estimator in a way that restricts the allowed $N$-point
functions of $\mathcal{L}(C_{\ell})$, but still captures most of the
relevant information. Explicitly, this is done through a specific
change of variables, such that the observed marginal posterior for
each multipole, $P(C_{\ell}|\mathbf{d})$, is transformed into a
Gaussian. Then, in these new variables the joint distribution is
approximated by a multivariate Gaussian. As long as the correlation
between any two multipoles is reasonably small, as is the case for
nearly full-sky experiments such as WMAP and Planck, we shall see that
this provides an excellent approximation to the exact joint
likelihood. As a result, the new approach greatly reduces the overall
number of samples required for convergence, and allows us to obtain a
highly accurate likelihood approximation to arbitrary
$\ell_{\textrm{max}}$. Generalization to a full polarized likelihood
will be discussed in a future paper (Eriksen et al., in preparation).

This paper is organized as follows: In \S \ref{sec:review}, we first
briefly review the Gibbs sampling algorithm together with the original
Blackwell-Rao estimator, and in \S\ref{sec:method} we introduce the
new Gaussianized Blackwell-Rao estimator. Next, in \S
\ref{sec:application}, we apply the new estimator to simulated data,
and compare results with brute-force likelihood evaluations in pixel
space. In \S \ref{sec:analysis}, we analyze the 5-year WMAP
temperature data, and provide an updated power spectrum and set of
cosmological parameters. We summarize and conclude in \S
\ref{sec:conclusion}. 

\section{Review of the CMB Gibbs sampler}
\label{sec:review}

We start by reviewing the current state of the CMB Gibbs sampling
framework, as previously developed through a series of papers
\citep{jewell:2004, wandelt:2004, eriksen:2004, larson:2007,
  eriksen:2008a}, and highlight the problem of likelihood modelling as
currently presented in the literature.

\subsection{Elementary CMB Gibbs sampling}

First, we assume that our observations, $\mathbf{d}$, in direction
$\hat{n}$ may be modelled in terms of a signal, $\mathbf{s}$ and a
noise, $\mathbf{n}$, component,
\begin{equation}
\mathbf{d}(\hat{n}) = \mathbf{s}(\hat{n}) + \mathbf{n}(\hat{n}).
\end{equation}
Further, we assume that both $\mathbf{s}$ and $\mathbf{n}$ are
Gaussian distributed with vanishing mean and covariances $\mathbf{S}$
and $\mathbf{N}$, respectively. The CMB is in this paper additionally
assumed to be isotropic, such that in spherical harmonic space
($\mathbf{s}(\hat{n}) = \sum_{\ell,m} a_{\ell m} Y_{\ell m}(\hat{n})$)
the CMB covariance matrix may be written as $\mathbf{S}_{\ell m,\ell'
  m'} = C_{\ell} \delta_{\ell \ell'} \delta_{mm'}$, where $C_{\ell} =
\left<a_{\ell m} a_{\ell m}^*\right>$ is the angular power
spectrum. Our goal is now to map out the CMB posterior distribution
$P(\mathbf{s}, C_{\ell}|\mathbf{d})$ and the CMB likelihood
$\mathcal{L}(C_{\ell}) = P(\mathbf{d}|C_{\ell})$. Note that we in this
paper are concerned with the problem of likelihood characterization
only, which is a post-processing step relative to the Gibbs
sampler. For notational transparency, we therefore neglect issues such
as foreground marginalization, instrumental beams, multi-frequency
observations etc. For full details on these issues, see, e.g., Eriksen
et al.\ 2008a.

When working with real-world CMB data, there are a number of issues
that complicate the analysis. Two important examples are anisotropic
noise and Galactic foregrounds. First, because of the scanning motion
of a CMB satellite, the pixels in a given data set are observed over
unequal amounts of time. This implies that the effective noise is a
function of pixel location on the sky. Second, large regions of the
sky are obscured by Galactic foregrounds (e.g., synchrotron, free-free
and dust emission), and these regions must be rejected from the
analysis by masking.

Because of such issues, the total data covariance matrix
$\mathbf{S}+\mathbf{N}$ is dense in both pixel and harmonic space. As
a result, it is computationally unfeasible to evaluate and sample
directly from $P(\mathbf{s}, C_{\ell}|\mathbf{d})$. Fortunately, this
problem was originally solved by \citet{jewell:2004},
\citet{wandelt:2004} and \citet{eriksen:2004}, who developed a
particular CMB Gibbs sampler for precisely this purpose. For full
details on this method, we refer the interested reader to the original
papers, and in the following we only describe the main ideas. The
practical implementation of the algorithm used in this paper is called
``Commander'', and has been described in detail by
\citet{eriksen:2004,eriksen:2008a}.

The idea behind the CMB Gibbs sampler is to draw samples from the
joint posterior by alternately sampling from the two corresponding
conditionals. The sampling scheme may thus be written on the symbolic
form
\begin{align}
\mathbf{s}^{i+1} &\leftarrow P(\mathbf{s}|C_{\ell}^{i}, \mathbf{d}) \\
C_{\ell}^{i+1} &\leftarrow P(C_{\ell} |\mathbf{s}^{i+1}, \mathbf{d}),
\end{align}
where the left arrow implies sampling from the distribution on the
right-hand side. Then, after some burn-in period, $(\mathbf{s}^{i},
C_{\ell}^{i})$ will be drawn from the desired
distribution. The only remaining step is to write down sampling
algorithms for each of the two above conditional distributions, both
of which are readily available for our problem, since the former is
simply a multivariate Gaussian, and the second is a product of
independent inverse Gamma distributions. For one possible general
sampling algorithm for $P(C_{\ell}|\mathbf{s})$, see, e.g.,
\citet{wehus:2008}.

\subsection{The Blackwell-Rao estimator}

The Gibbs sampler produces a set of samples drawn from the joint CMB
posterior, $P(\mathbf{s}, C_{\ell}|\mathbf{d})$. However, for these
samples to be useful for estimation of cosmological parameters, we
have to transform the information contained in this sample set into a
smooth approximation to the likelihood $\mathcal{L}(C_{\ell}) =
P(\mathbf{d}|C_{\ell})$. In principle, we could simply generate a
multi-variate histogram and read off corresponding values, but this
does not work in practice because of the large dimensionality of the
parameter space.

In the current literature, the best approach for handling this problem
is the Blackwell-Rao (BR) estimator \citep{wandelt:2004,chu:2005},
which attempts to smooth the sampled histogram by taking advantage of
the known analytic distribution, $P(C_{\ell}|\mathbf{s})$: First, we
define the observed power spectrum, $\sigma_{\ell}$, of the current
CMB sky Gibbs sample, $\mathbf{s}(\hat{n}) = \sum_{\ell,m} a_{\ell m}
Y_{\ell m}(\hat{n})$,
\begin{equation}
  \sigma_\ell \equiv \frac{1}{2\ell +1}\sum_{m=-\ell}^\ell |a_{\ell m}|^2.
\end{equation}
Then the BR estimator is derived as follows,
\begin{align}
P(C_{\ell}|\mathbf{d}) &= \int P(C_{\ell}, \mathbf{s}|\mathbf{d})
\,d\mathbf{s}\notag\\ 
&= \int P(C_{\ell}|\mathbf{s},\mathbf{d})
P(\mathbf{s}|\mathbf{d})\,d\mathbf{s} \notag\\ 
&= \int P(C_{\ell}|\sigma_{\ell})
P(\sigma_{\ell}|\mathbf{d})\,D\sigma_{\ell}\notag\\ &
\approx \frac{1}{N_{\textrm{G}}}\sum_{i=1}^{N_{\textrm{G}}}
P(C_{\ell}|\sigma_{\ell}^{i}).
\label{eq:br}
\end{align}
In other words, the BR estimator is nothing but the average of
$P(C_{\ell}|\sigma_{\ell})$ over the sample set, where $\sigma_{\ell}$
refers to the power spectrum of a full-sky noiseless CMB signal Gibbs
sample. This distribution has a simple analytic expression
\citep[e.g.,][]{chu:2005},
\begin{equation}
  P(C_\ell | \mathbf \sigma_\ell) \propto \prod_{\ell}
  \frac{e^{-\frac{2\ell+1}{2} \frac{\sigma_{\ell}}{C_{\ell}}}}
{C_{\ell}^{\frac{2\ell+1}{2}}}.
\label{eq:prob_sigma}
\end{equation}

While Equation \ref{eq:br} does constitute a computationally
convenient and accurate approximation to the full likelihood for some
special applications, it suffers badly from poor convergence
properties with increasing dimensionality of the sampled space. This
behaviour may be understood in terms of relative distribution widths:
Suppose we want to map out an $\ell_{\textrm{max}}$-dimensional
distribution, and each of the univariate Blackwell-Rao functions
[i.e., $P(C_{\ell}|\sigma_{\ell})$] have a standard deviation of, say,
90\% of the corresponding marginal distributions. The total volume
fraction spanned by a single sample in $\ell_{\textrm{max}}$
dimensions is then $f = 0.9^{\ell_{\textrm{max}}}$, an exponentially
decreasing function with $\ell_{\textrm{max}}$. Therefore it also
takes an exponential number of samples in order to build up the full
histogram, and this becomes computationally unfeasible for realistic
data sets already at $\ell_{\textrm{max}} \gtrsim 30-50$ \citep{chu:2005}.

\begin{figure*}
\mbox{\epsfig{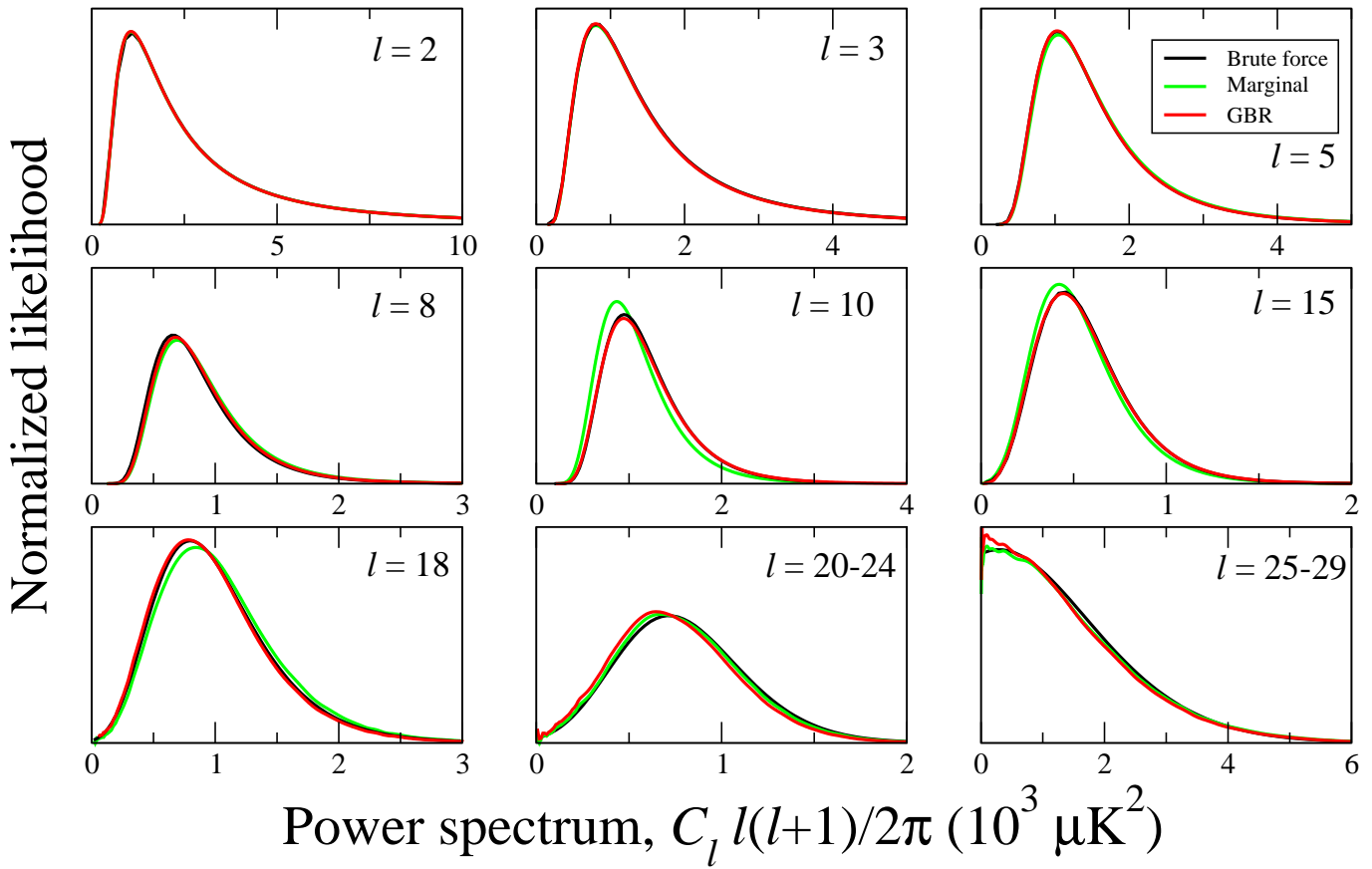}}
\caption{Slices through the joint likelihood from a low-resolution
  simulation, computed by brute-force pixel space evaluation (black
  lines) and the Gaussianized Blackwell-Rao estimator (red
  lines). Green lines show the marginal distribution for each $\ell$,
  to illustrate the effect of mode coupling caused by the WMAP KQ85
  sky cut used in this analysis. }
\label{fig:verification}
\end{figure*}

The main problem with this approach is that one attempts to map out
all possible $N$-point correlation functions between all
multipoles. The number of such $N$-point functions is obviously
overwhelming with increasing dimensionality. But this also hints at a
possible resolution of the problem: We know by experience that the CMB
likelihood is a reasonably well behaved function, in that 1) there are
only weak correlations between multipoles for data sets with nearly
full-sky coverage, and 2) that even including just two-point
correlations (in transformed variables) produces very reasonable
results (e.g., Bond, Jaffe \& Knox 2000; Verde et al.\ 2003). This
intuition will be used in the next section to define a stable
likelihood estimator.

\section{The Gaussianized Blackwell-Rao estimator}
\label{sec:method}

We now introduce a new Gibbs-based likelihood estimator we call the
``Gaussianized Blackwell-Rao estimator'' (GBR). The basic idea behind
this approach is similar to that employed by, e.g., \citet{bond:2000}
and \citet{hamimeche:2008}, namely approximation by a multivariate
Gaussian in a transformed set of variables.

Explicitly, our approximation is defined by transforming the
univariate marginal distributions $P(C_{\ell}|\mathbf{d})$ into
Gaussianized variables, $x_{\ell}$, and then assuming a multivariate
Gaussian distribution in these transformed variables,
\begin{equation}
  P(C_\ell | \mathbf d) = \left(\prod_{\ell}\frac{\partial
      C_\ell}{\partial x_{\ell}}\right)^{-1} P(\mathbf{x} | \mathbf
  d). 
\label{eq:transformation}
\end{equation}
Here $\frac{\partial C_\ell}{\partial x_{\ell}}$ is the Jacobian of
the transformation, and $\mathbf{x} = \{x_{\ell}\}$ is a Gaussian
random vector with mean $\mu = \{\mu_{\ell}\}$ and covariance matrix
$\mathbf{C}_{\ell \ell'} = \left
  <(x_{\ell}-\mu_{\ell})(x_{\ell'}-\mu_{\ell'})\right>$. Thus, the
approximation of our likelihood estimator relies on the assumption
that
\begin{equation}
  P(\mathbf{x}|\mathbf{d}) \approx e^{-\frac{1}{2}(\mathbf{x}-\mu)^T \mathbf{C}^{-1} (\mathbf{x}-\mu)}.
\end{equation}
Note that this is by construction exact for the full-sky uniform noise
case, because the covariance matrix in this case is diagonal, and the
full expression factorizes in $\ell$; in that case we are only
performing an identity operation. 

\subsection{Transformation to Gaussian marginal variables}

The first step in our algorithm is to compute the change-of-variables
rule from $C_{\ell}$ to $x_{\ell}$ that transforms the marginal
distribution, $P(C_{\ell}|\mathbf{d})$, for each $\ell$ into a
Gaussian distribution, $P(x_{\ell}|\mathbf{d})$. The data used for
this process are the $\sigma_{\ell}$ samples drawn from the joint
posterior $P(C_{\ell}, \mathbf{s}|\mathbf{d})$ by the CMB Gibbs
sampler.

We use two different methods of estimating the marginal distributions
from these samples. The first approach is to estimate
$P(C_{\ell}|\mathbf{d})$ with the Blackwell-Rao estimator as defined
by Equation \ref{eq:br}, over a grid in $C_{\ell}$ for each
$\ell$. Then, a cubic spline is fitted to the resulting distribution.
This is the preferred approach for high signal-to-noise or low-$\ell$
modes.

However, for low signal-to-noise and high-$\ell$ modes one observes
similarly poor convergence properties of this marginal estimator as
for the full joint estimator. In these cases we therefore instead
compute a simple histogram directly from the $C_{\ell}$ samples, and
fit a smooth spline \citep{green:1994} through this histogram. For
further stability, we also produce $\mathcal{O}(10^6)$ $C_{\ell}$
samples from $P(C_{\ell}|\sigma_{\ell})$ based on the same
$\sigma_{\ell}$ set as used for the BR estimator. This essentially
corresponds to computing the Blackwell-Rao estimator by Monte Carlo,
and the computational cost of producing these extra samples is
small. (The computational expense of the Gibbs sampler is driven by
sampling from $P(\mathbf{s}|C_{\ell},\mathbf{d})$, not by
$P(C_{\ell}|\mathbf{s})$.)  Note that this approach naturally supports
arbitrary $C_{\ell}$ binning schemes \citep{wehus:2008}, and also
interfaces naturally with the hybrid MCMC scheme described by
\citet{jewell:2004}.

Given these spline approximations to $P(C_{\ell}|\mathbf{d})$ for each
$\ell$, we compute the corresponding cumulative distributions by
numerical integration,
\[
F(C_\ell | \mathbf{d}) = \int_{0}^{C_\ell} P(C_\ell' | \mathbf{d}) dC_\ell'.
\]
This is subsequently identified with a standard Gaussian distribution
with zero mean and unity variance. Explicitly, we find
$x_\ell(C_\ell)$ over a grid in $C_\ell$ such that
\[
F(C_\ell|\mathbf{d}) = F_{\textrm{Gauss}}(x_\ell) = \frac{1}{2}\left(1+\textrm{erf}\left(\frac{x_{\ell}}{\sqrt{2}}\right)\right),
\]
where $\textrm{erf}$ is the error function. This equation is
straightforward to solve using standard numerical root-finding
routines. The result is a convenient set of look-up tables
$x_{\ell}(C_{\ell})$, again stored in the form of cubic splines, that
allows for very efficient transformation from standard to Gaussian
variables for arbitrary values of $C_{\ell}$. From these splines, it
is also easy to compute the derivatives required for the Jacobian in
Equation \ref{eq:transformation}.

\subsection{Estimation of the joint Gaussian density}

Having defined a change-of-variables for each $\ell$, the remaining
task is to estimate the joint distribution,
$P(\mathbf{x}|\mathbf{d})$, in the new variables. In this paper, we
approximate this distribution by a joint Gaussian, but any parametric
function could of course serve this purpose. For example, we
implemented support for the skew-Gaussian distribution
\citep[e.g.,][]{azzalini:2003} in our codes, but found that the
improvement over a simple Gaussian was very small.

The only free parameters in this multivariate Gaussian distribution
are the mean, $\mu$, and the covariance, $\mathbf{C}$. These are again
estimated from the samples produced by the Gibbs sampler. First, we
draw $N \sim \mathcal{O}(10^6)$ $C_{\ell}$ samples from
$P(C_{\ell}|\sigma_{\ell})$, as described above, but this time
including all $\ell$'s for each sample. Then we Gaussianize these
$\ell$-by-$\ell$, by evaluating $x_{\ell}(C_{\ell})$ for each sample
and multipole moment. Finally, we compute the corresponding means and
standard deviations,
\begin{align}
\mu_{\ell} &= \frac{1}{N}\sum_{i=1}^{N} x_{\ell}^i \\
C_{\ell\ell'} &= \frac{1}{N}\sum_{i=1}^{N}
(x_{\ell}^i-\mu_{\ell})(x_{\ell'}^i - \mu_{\ell'}),
\end{align}
where the sums run over sample index. 

\section{Application to simulated data}
\label{sec:application}

Before applying the machinery described in the previous section to the
5-year WMAP data, we verify the method by a analyzing a simulated
low-resolution data set. The reason for considering a low-resolution
simulation is that only in this case is it possible to evaluate the
exact likelihood by brute force in pixel space, without making any
approximations.

The simulation is made by drawing a Gaussian realization from the
best-fit 5-year WMAP $\Lambda$CDM power spectrum \citep{komatsu:2008},
smoothing this with a $10^{\circ}$ FWHM Gaussian beam, and projecting
it on an $N_{\textrm{side}}=16$
HEALPix\footnote{http://healpix.jpl.nasa.gov} grid. Finally,
$20\mu\textrm{K}$ RMS white noise is added to each pixel, and the
(degraded) WMAP KQ85 sky cut \citep{gold:2008} is applied to the
data. The maximum multipole considered in this analysis was
$\ell_{\textrm{max}} = 47$, and the spectrum was binned with a bin
size of $\Delta \ell=5$ from $\ell=20$. The signal-to-noise is unity
at $\ell=19$, and negligible beyond $\ell \ge 30$.

\begin{figure}
\mbox{\epsfig{file=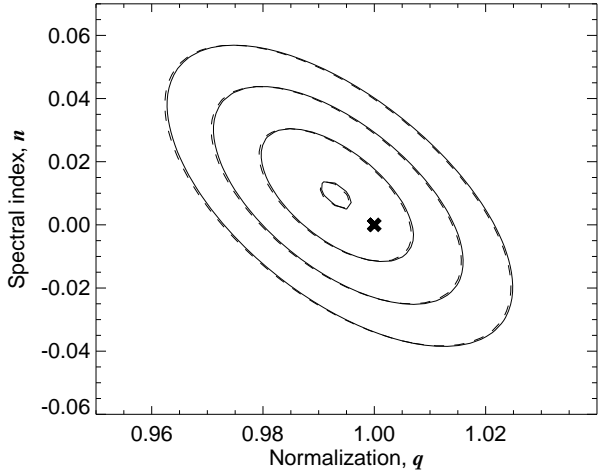,width=\linewidth,clip=}}
\caption{Comparison of amplitude---tilt likelihoods derived from the
  5-year WMAP GBR estimator up to $\ell \le 250$, for two independent
  sample sets (solid and dashed lines). Contours are where
  $-2\textrm{ln}\mathcal{L}(C_{\ell})$ rises by 0.1, 2.3, 6.17 and
  11.8 from the minimum, corresponding to the peak and 1, 2 and
  $3\sigma$ confidence regions in a Gaussian distribution. See main
  text for full details. The cross marks the point $(q,n) = (1,0)$,
  corresponding to the best-fit model for WMAP including all $\ell$'s,
  and this is found to lie well inside the $1\sigma$ contour.}
\label{fig:qn_model}
\end{figure}

We now compute slices for each $\ell$ through the full multivariate
likelihood, both with the method described in \S\ref{sec:method} and
by brute-force pixel space evaluation \citep[e.g.,][]{eriksen:2007a},
fixing all other $\ell$'s at the input $\Lambda$CDM spectrum. For
comparison, we also compute the the marginal distributions for each
$\ell$.

The results from this exercise are shown in Figure
\ref{fig:verification}. Black lines indicate the brute-force
likelihoods, and red lines show the Gaussianized Blackwell-Rao
likelihoods. The green lines show the marginal distributions,
visualizing the effect of mode coupling due to the sky cut.

First, we see that all distributions agree very closely at
$\ell\le8$. In this very large-scale regime, all harmonic modes are
sufficiently well sampled with the KQ85 sky cut that mode coupling is
negligible. However, from $\ell\ge10$ the marginal distributions are
noticeably different from the likelihood slices, with a typical shift
in peak position of $\sim100\mu\textrm{K}^2$'s. We also see that these
correlations are accurately captured by the Gaussian approximation
implemented in the GBR estimator, as the GBR likelihoods are
essentially identical to the brute-force slices up to $\ell = 18$.

At the very high $\ell$ and low signal-to-noise end, we see slight
differences between the GBR and the pixel space slices, and in fact,
the agreement is better with the marginal distributions. This is
caused by poor convergence of the covariance matrix in this particular
run, and is included here for pedagogical purposes only: In a real
analysis, one must always make sure that all distributions have
converged well, typically by analyzing different chain sets
separately. Note also that with sufficiently wide bins, the
correlations to neighboring bins eventually vanish, and in this case
it may be better to remove these correlations by hand from the
covariance matrix, rather than trying to estimate them by
sampling. Whether this is the case or not for a given set can again be
estimated by jack-knife tests. Finally, for the 5-year WMAP analysis
presented in this paper, we will only use the GBR estimator in the
high signal-to-noise regime, and in that case the distributions
converge very quickly.

\section{5-year WMAP temperature analysis}
\label{sec:analysis}

We now apply the tools described in \S\ref{sec:method} to the 5-year
WMAP temperature data. We only consider $\ell \le 200$ in this paper,
to avoid issues with error propagation for unresolved point sources
and beam estimation. However, we do correct for the mean spectrum of
unresolved point sources, as described below.

\subsection{Data}

We analyze the foreground reduced 5-year WMAP V-band temperature sky
maps, which are available from
Lambda\footnote{http://lambda.gsfc.nasa.gov}. The V-band data was
chosen because these are considered to be the cleanest in terms of
foregrounds out of the five WMAP bands \citep{gold:2008}. Further, at
$\ell \le 200$ the V-band alone is strongly cosmic variance dominated,
and one does therefore not gain any significant statistical power by
co-adding with other bands. Instead, one only increases the chance of
introducing foreground biases by adding more frequencies. We work with
the individual differencing assembly (DA) maps \citep{hinshaw:2003},
and take into account the beam and noise pattern for each map
separately.

\begin{figure}
\mbox{\epsfig{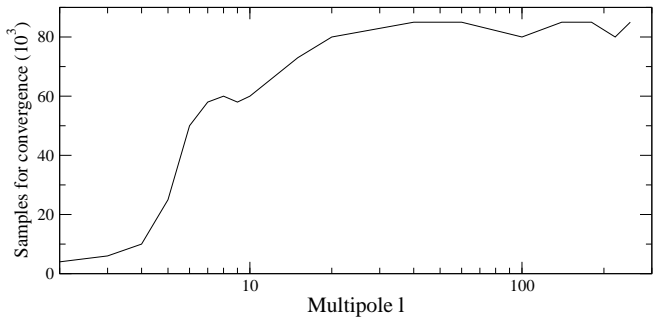}}
\caption{Number of $C_{\ell}$ samples required for the GBR covariance
  matrix to converge, as a function of $\ell_{\textrm{max}}$.}
\label{fig:covar_convergence}
\end{figure}

\begin{figure}
\mbox{\epsfig{file=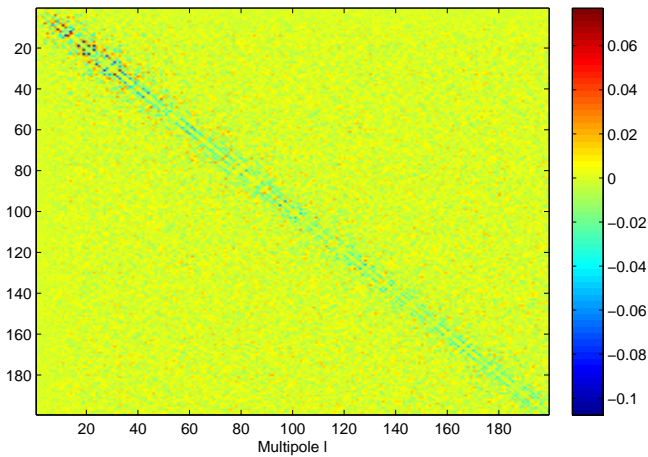,width=\linewidth,clip=}}
\caption{GBR correlation matrix, $\tilde{\mathbf{C}}_{\ell\ell'}$, for
  the 5-year WMAP V-band data.}
\label{fig:covar}
\end{figure}

The WMAP sky maps are pixelized at a HEALPix resolution of
$N_{\textrm{side}}=512$, corresponding to a pixel size of $7'$, and
the instrumental beam of the two V-band channels has a FWHM of 21'. We
therefore impose an upper harmonic mode limit of
$\ell_{\textrm{max}}=700$ in the Gibbs sampling (Commander) step,
probing deeply into the noise dominated regime. Note, however, that we
only use $\ell \le 200$ in the GBR estimator, to avoid high-$\ell$
complications, such as beam and point source error propagation, in the
cosmological parameter stage.

We correct the spectrum for unresolved point sources using the WMAP
model. Explicitly, the mean spectrum due to unresolved point sources
in a single frequency, $\nu$, for the 5-year WMAP data is modelled as
\citep{hinshaw:2003,hinshaw:2007,nolta:2008}
\begin{equation}
C_{\ell}^{\textrm{ps}} = A_{\textrm{ps}} a(\nu)^2 \left(\frac{\nu}{\nu_0}\right)^{2\beta},
\end{equation}
where $A_{\textrm{ps}} = 0.011\pm0.001$ is the point source amplitude
relative to the Q-band channel ($\nu_0 = 41\textrm{GHz}$),
$\beta=-2.1$ is the best-fit spectral index of the point sources, and
$a(\nu)$ is the conversion factor between antenna and thermodynamic
temperature units. To correct for this in our analysis, we subtract
$C_{\ell}^{\textrm{ps}}$, evaluated at $\nu=61\textrm{GHz}$, from each
$\sigma_{\ell}$ sample before computing the GBR estimator.  

Finally, we impose the WMAP KQ85 sky cut \citep{gold:2008} on the data
that masks point sources, removing 18\% of the sky. Note that we adopt
the template corrected maps provided by the WMAP team in this
analysis, and postpone an internal Gibbs sampling based foreground
analysis to a future paper; for now our main focus is the new
likelihood approximation, not the impact of foregrounds.

\begin{figure}
\mbox{\epsfig{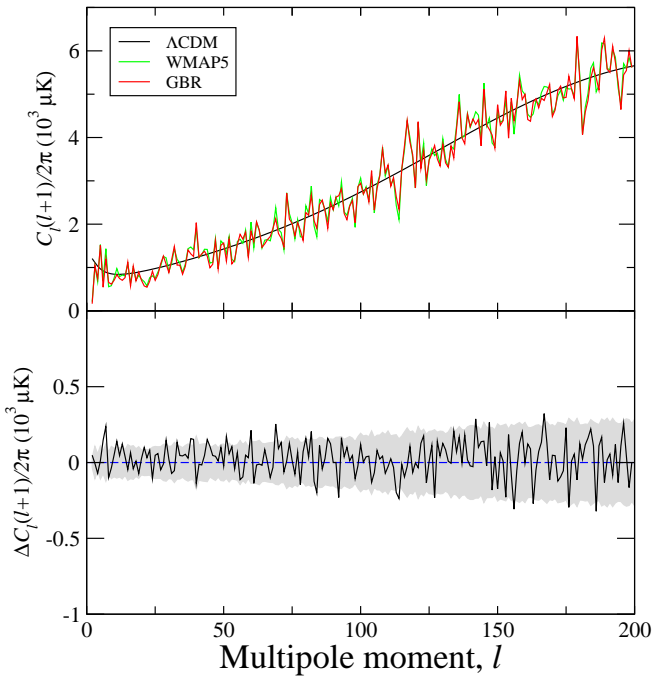}}
\caption{Comparison of 5-year power spectra obtained by WMAP and
  Commander/GBR up to $\ell=200$. The bottom panel shows the
  difference of these two spectra, and the gray band indicates the
  68\% confidence region due to noise and sky cut only, not cosmic
  variance.}
\label{fig:powspec}
\end{figure}

\subsection{Analysis overview}

The analysis consists of the following steps:
\begin{enumerate}
\item Generate 4000 $\sigma_{\ell}$ samples with Commander from the
  5-year V1 and V2 differencing assemblies , including $\ell$'s up to
  $\ell_{\textrm{max}}=700$, divided over 8 chains.
\item Generate 500\,000 $C_{\ell}$ samples from these $\sigma_{\ell}$'s,
  including $\ell$'s between $\ell=2$ and 250.
\item Compute the corresponding GBR parameters, i.e., transformation
  tables, means $\mu$ and covariance matrix $\mathbf{C}$.
\item Modify the 5-year WMAP temperature likelihood by replacing the
  existing low-$\ell$ part with Equation \ref{eq:transformation}, with
  the parameters given in (3). The transition multipole between the
  low-$\ell$ and high-$\ell$ is increased from $\ell=32$ to
  200. Multipoles between $\ell=201$ and 250 are included in the GBR
  estimator to avoid truncation effects, but the spectrum in this
  range is kept fixed at a fiducial spectrum, in order not to count
  these multipoles twice.
\item Cosmological parameters are estimated using CosmoMC \citep{lewis:2002}.
\end{enumerate}

\begin{figure*}
\mbox{\epsfig{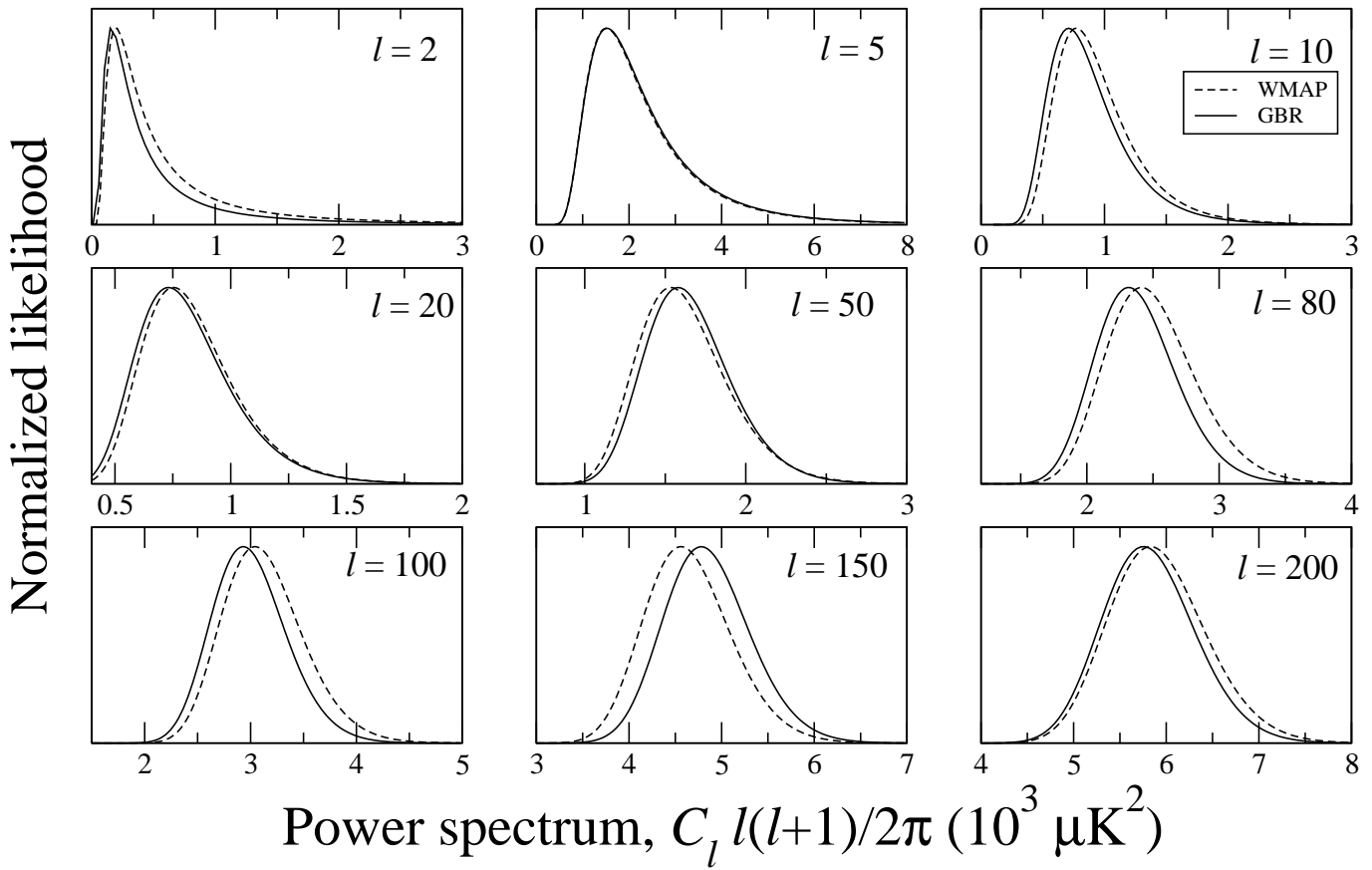}}
\caption{Comparison of likelihood slices from the standard WMAP
  likelihood code (dashed lines) and the new GBR likelihood (solid
  lines).}
\label{fig:likelihood_comparison}
\end{figure*}

\subsection{Convergence analysis}

Before presenting the results from the WMAP analysis, we consider the
question of convergence. First, we compute the Gelman-Rubin statistic
\citep{gelman:1992} for each $\sigma_{\ell}$ using the eight chains
computed with Commander and removing the first 20 samples for
burn-in. We find that $R-1$ is less than 0.01 for $\ell \lesssim 300$
and less than 1.1 for $\ell \lesssim 500$, indicating very good
convergence in terms of power spectra.

However, the fact that each $\sigma_{\ell}$ individually is well
converged does not automatically imply that the full likelihood is
well converged, since the latter depends crucially on the correlations
between $\sigma_{\ell}$'s. To assess the convergence in terms of
cosmological parameters, we therefore analyse a toy model, by fitting
a simple two-parameter amplitude and tilt, $q$ and $n$, model,
\begin{equation}
C_{\ell} = q \,\left(\frac{\ell}{\ell_{\textrm{pivot}}}\right)^{n}\,C_{\ell}^{\textrm{fid}},
\end{equation}
to the WMAP data between $\ell=2$ and 250 with the GBR
likelihood. Here $C_{\ell}^{\textrm{fid}}$ is a fiducial power
spectrum, which is chosen to be the best-fit 5-year WMAP $\Lambda$CDM
power spectrum \citep{komatsu:2008}, and
$\ell_{\textrm{pivot}}=150$. We then map out the likelihood in a grid
over $q$ and $n$. This is repeated twice, first including samples from
chains number 1 to 4 and then from chains number 5 to 8. 

The results from this exercise are shown in Figure \ref{fig:qn_model}
in terms of two sets of likelihood contours, corresponding to each of
the two chain sets, respectively. The agreement between the two is
excellent, indicating that we also have good convergence in terms of
cosmological parameters with the existing sample set. Note also that
the point $(q,n) = (1,0)$ lies well inside the $1\sigma$ confidence
region, indicating that the best-fit WMAP model, which is obtained
including $\ell$'s between $\ell=2$ and 1024, also is a good fit to
$\ell=2$ to 250 separately.

Third, as described in \S\ref{sec:method}, we construct the GBR
covariance matrix from $N=\mathcal{O}(10^6)$ $C_{\ell}$ samples drawn
from the (smaller) set of $\sigma_{\ell}$ samples. An outstanding
question is how large $N$ should be in order for this covariance
matrix to reach convergence, as a function of
$\ell_{\textrm{max}}$. To settle this question, we carry out the
following simple exercise: First we produce two $C_{\ell}$ sample
sets, each containing $N$ samples, and all drawn from a single
$\sigma_{\ell}$ sample. Second, we compute the two corresponding
covariance matrices, invert these, then subtract them from each other,
and finally compute the standard deviation of all elements. Third, we
define the inverse covariance matrix to be converged if the RMS MCMC
noise is less than 0.005, corresponding to 0.5\% of the diagonal
elements. (We have checked that this produces robust parameter
estimates.) We then find the smallest $N$ such that this is satisfied,
as a function of $\ell_{\textrm{max}}$.

The results from this exercise are shown in Figure
\ref{fig:covar_convergence}. Here we see that the number of samples
required for convergence rises rapidly up to $\ell\sim30$, reaching a
maximum of $\sim8\times10^4$ samples, and then flattens to a plateau. To be on
the safe side, we therefore always use either $5\times10^5$ or $10^6$
samples in the WMAP analysis.

The reason for this behaviour becomes intuitive when considering the
structure of the actual matrix. This is shown in Figure
\ref{fig:covar}, on the form of a correlation matrix
\begin{equation}
\tilde{C}_{\ell\ell'} =
\frac{C_{\ell\ell'}}{\sqrt{C_{\ell\ell}C_{\ell'\ell'}}}-\delta_{\ell\ell'}.
\end{equation}
The main features of this matrix are negative correlations around the
diagonal, with the largest amplitudes observed between $\ell$ and
$\ell\pm2$. This is expected: First, two modes separated by
$\Delta\ell=1$ have different parity, and can therefore not easily
mimic each other. On the other hand, modes separated by $\Delta\ell=2$
have both identical parity and similar angular scale, and it is
therefore possible to add power to one mode and subtract it from the
other, and still maintain an essentially unchanged image. The result
is a noticeable anti-correlation between $\ell$ and $\ell\pm2$.

At larger separations in $\ell$, the correlations die off rapidly,
since it is difficult for a large-scale mode to mimic a small-scale
mode with a reasonably small sky cut. And this explains the
convergence behaviour seen in Figure \ref{fig:covar_convergence}: The
covariance matrix is strongly band-limited. Therefore, once one has a
sufficiently large number of $C_{\ell}$ samples for a sub-block to
converge, there is also enough samples for a sub-block further away to
converge. These are essentially uncorrelated. 

\subsection{Results}

We now present the main results derived from the 5-year WMAP
temperature data with the GBR estimator between $\ell=2$ and 200.
First, in the top panel of Figure \ref{fig:powspec} we plot the power
spectrum obtained by maximizing the GBR likelihood together with the
official 5-year WMAP power spectrum. The bottom panel shows the
difference between these two, and the gray band indicates the standard
deviation of $\sigma_{\ell}$, i.e., the uncertainty due to noise and
sky cut, but not to cosmic variance. Clearly, the agreement between
the two power spectra is very good.

\begin{figure*}
\mbox{\epsfig{file=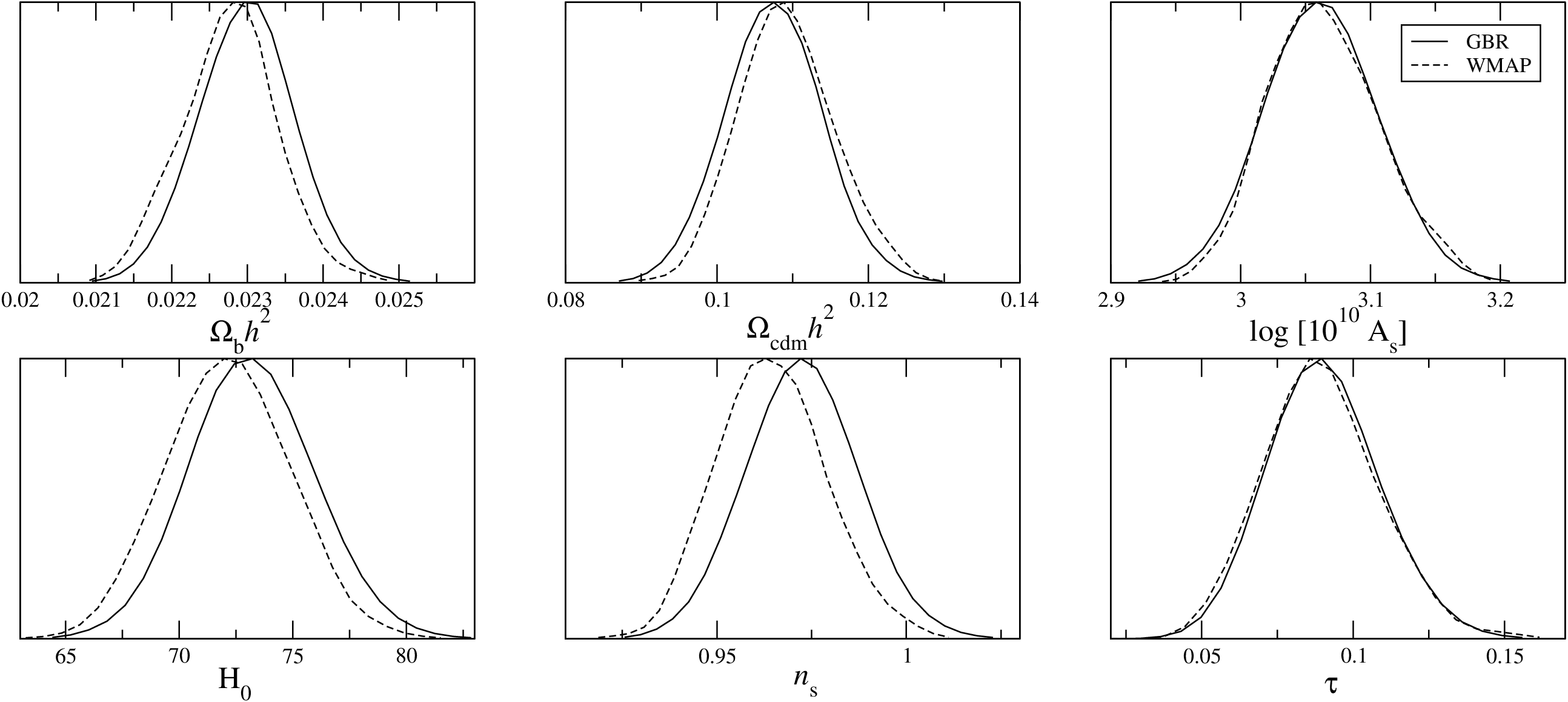,width=\linewidth,clip=}}
\caption{Comparison of marginal cosmological parameter posteriors
  obtained with the standard WMAP likelihood code (dashed lines) and
  with the modified GBR likelihood code up to $\ell=200$ (solid line)
  for 5-year WMAP data.}
\label{fig:parameters}
\end{figure*}

Next, in Figure \ref{fig:likelihood_comparison} we compare a few
selected slices through the GBR likelihood with slices through the
WMAP likelihood. All other $\ell$'s than the one currently considered
are kept fixed at the best-fit $\Lambda$CDM spectrum. Here we see
that there are small shifts in peak positions, corresponding to the
small differences seen in the power spectra in Figure
\ref{fig:powspec}. However, a main point in this plot is that the GBR
likelihood slices are well behaved even at the highest $\ell$'s, and
this is not the case for the standard BR estimator
\citep{chu:2005}. 

Finally, in Table \ref{tab:parameters} and Figure \ref{fig:parameters}
we summarize the marginal cosmological parameter posteriors obtained
with the two likelihood codes from CosmoMC. Interestingly, there are
some notable differences at the 0.3--$0.6\sigma$ level, with the most
striking example being the spectral index of scalar perturbations,
$n_{\textrm{s}}=0.973\pm0.014$. This is only $1.9\sigma$ away from
unity, and $0.6\sigma$ higher than the official WMAP values. 

\section{Conclusions}
\label{sec:conclusion}

We have presented a new likelihood approximation to be used within the
CMB Gibbs sampling framework. This approximation is defined by
Gaussianizing the observed marginal power spectrum posteriors,
$P(C_{\ell}|\mathbf{d})$, through a specific change-of-variables, and
then coupling these univariate posteriors into a joint distribution
through a multivariate Gaussian in the new variables. This process is
exact, i.e., an identity operation, in the uniform and full-sky
coverage case, and it is also an excellent approximation in for the
moderate sky cuts relevant to satellite missions such as WMAP and
Planck.

Our new approach relies on the previously described CMB Gibbs sampling
framework \citep{jewell:2004,wandelt:2004,eriksen:2004}, and thereby
inherits many important advantages from that. First and foremost, this
framework allows for seamless propagation of uncertainties from
various systematic effects (e.g., foregrounds, beam uncertainties,
calibration or noise estimation errors) to the final cosmological
parameters. This is not straightforward in the hybrid scheme used by
the WMAP code. Second, this new approach corresponds to the exact
low-$\ell$ pixel space likelihood part of the WMAP code, not the
approximate high-$\ell$ MASTER part. Still, our method can handle
arbitrary high $\ell$'s. Third, once the one-time pre-processing step
has been completed, the computational expense of our estimator is
determined by the cost of $\ell_{\textrm{max}}$ spline evaluations,
while a pixel space approach requires a matrix inversion, and
therefore scales as $\mathcal{O}(N_{\textrm{pix}}^3)$. For the cases
considered in this paper, the CPU time required for the GBR WMAP
estimator up to $\ell=200$ was $\sim0.2$ milliseconds, while it was
$\sim20$ seconds for the pixel space approach up to $\ell=32$, for a
map with $~2500$ pixels.

In order to validate our estimator, we applied it to a low-resolution
simulated data set, and compared it to slices through the exact joint
likelihood as computed by brute-force evaluation in pixel space. The
agreement between the two approaches was excellent. We then applied
the same estimator to the 5-year WMAP temperature data, and estimated
both a new power spectrum and new cosmological parameters within a
standard six-parameter $\Lambda$CDM model.

The results from these calculations are interesting. First, our power
spectrum is statistically very similar to the official WMAP spectrum,
with no visible biases seen and relative fluctuations within the level
predicted by noise and sky cut. Nevertheless, we do find significant
differences in terms of cosmological parameters, and most notably in
the spectral index of scalar perturbations,
$n_{\textrm{s}}$. Specifically, we find $n_{\textrm{s}} = 0.973\pm
0.014$, which is only $1.9\sigma$ away from unity and $0.6\sigma$
higher than the official WMAP result, $n_{\textrm{s}} = 0.965\pm
0.014$.

This result resembles very much the outcome of a re-analysis we did
with the 3-year WMAP temperature data \citep{eriksen:2007a}, for which
we found a bias of $0.4\sigma$ in $n_{\textrm{s}}$ compared to the
official WMAP results. This bias was due to the sub-optimal
MASTER-based likelihood approximation \citep{hivon:2002,verde:2003}
used by the WMAP team between $\ell=12$ and 30, whereas we used an
exact estimator in the same range. This study later prompted the WMAP
to change their codes to use an exact likelihood evaluator up to
$\ell=30$.

In the same study, we also tried to increase the $\ell$-range for our
exact estimator to $\ell=50$, but found small differences. We
therefore concluded, perhaps somewhat prematurely, that an exact
estimator up to $\ell=30$ was sufficient for obtaining accurate
results. On the contrary, in this paper we find still find significant
changes when increasing the exact estimator up to $\ell=200$.

In retrospect, this should perhaps not come as a complete surprise,
when realizing that the impact on a particular cosmological parameter
typically depends logarithmically on $\ell$. For instance,
\citet{hamimeche:2008} considered a simple power spectrum model with a
single free amplitude, $C_{\ell} = q\,C_{\ell}^{\textrm{fid}}$, and
found that, for a given likelihood estimator to be ``statistically
unbiased'', the systematic errors in that same estimator must fall off
faster than $\sim 1/\ell$.

\begin{deluxetable}{cccc}
\tablewidth{0pt}
\tablecaption{Marginal 5-year WMAP cosmological parameters\label{tab:parameters}} 
\tablecomments{Comparison of cosmological parameters obtained with the
standard 5-year WMAP likelihood code (second column) and with the new
GBR estimator at $\ell\le200$ (third column), given in terms of marginal
means and standard deviations. The shift between the two in units of
$\sigma$ is listed in the fourth column.}
\tablecolumns{4}
\tablehead{Parameter  & WMAP  & GBR  & Shift in $\sigma$  
}
\startdata
$\Omega_{\textrm{b}}h^2$ & $0.0228\pm0.0006$ & $0.0230\pm0.0006$ &    0.4 \\
$\Omega_{\textrm{c}}h^2$ & $0.109\pm0.006$ & $0.0108\pm0.006$ &    -0.3 \\
$\textrm{log}(10^{10}A_{\textrm{s}})$  & $3.06\pm0.04$ & $3.06\pm0.04$ &    0.0 \\
$h$                    & $0.722\pm0.03$ & $0.732\pm0.03$ &     0.3 \\
$n_{\textrm{s}}$         & $0.965\pm0.014$ & $0.973\pm0.014$ &    0.6 \\
$\tau$                 & $0.090\pm0.02$ & $0.090\pm0.02$ &    0.0
\enddata
\end{deluxetable}

A similar consideration holds for $n_{\textrm{s}}$. Intuitively,
$n_{\textrm{s}}$ is as much affected by $\ell=2$ to 10 as it is
between $\ell=20$ and 100. In the previous 3-year WMAP re-analysis
paper, we increased the range of the accurate likelihood estimator
from $\ell=12$ to 30, corresponding to a factor of 2.5 in $\ell$, and
removed a bias of $\sim0.4\sigma$ in $n_{\textrm{s}}$. In this paper,
we increase the range from $\ell=30$ to 200, corresponding to a factor
of 6.7 in $\ell$, and find an additional bias of $0.6\sigma$. However,
increasing $\ell$ from 30 to 50 corresponds only to a factor of 1.7 in
$\ell$, and this appears to be too small to produce a statistically
significant result.

The main conclusions from this work are two-fold. First, it seems that
an accurate likelihood description is required to higher $\ell$'s than
previously believed, and at least up to $\ell=200$, in order to obtain
unbiased results. By extrapolation, it also does not seem unlikely
that even higher multipoles should be included. This issue will be
revisited in a future publication. 

Our second main conclusion is that we find a spectral index only
$1.9\sigma$ away from unity, namely $n_{\textrm{s}} =
0.973\pm0.014$. To us, it therefore seem premature to make strong
claims concerning $n_{\textrm{s}}\ne 1$; the statistical significance
of this is rather low, and there are likely still unknown systematic
errors in this number.

In a future publication we will generalize the GBR estimator to
polarization. Once completed, this will enable a fully Gibbs-based CMB
likelihood analysis at low $\ell$'s, and remove the need for
likelihood techniques based on matrix operations, i.e., inversion and
determinant evaluation. The computational cost of a standard
cosmological parameter MCMC analysis (e.g., CosmoMC) will then once
again be driven by the required Boltzmann codes (e.g., CAMB or
CMBFast) and not by the likelihood evaluation. In turn, this will
increase the importance of fast interpolation codes such as Pico
\citep{fendt:2007} or COSMONET \citep{auld:2007}. With such fast
algorithms for both spectrum and likelihood evaluations ready at hand,
the CPU requirements for cosmological parameter estimation may
possibly be reduced by orders of magnitude.

\begin{acknowledgements}
  We thank Tony Banday, Ben Wandelt and Graca Rocha for useful and
  interesting discussions. We acknowledge use of the HEALPix software
  \citep{gorski:2005} and analysis package for deriving the results in
  this paper. We acknowledge use of the Legacy Archive for Microwave
  Background Data Analysis (LAMBDA). This work was partially performed
  at the Jet Propulsion Laboratory, California Institute of
  Technology, under a contract with the National Aeronautics and Space
  Administration. {\O}R, NEG and HKE acknowledge financial support
  from the Research Council of Norway.
\end{acknowledgements}


\begin{thebibliography}{}

\bibitem[Auld et al.(2007)]{auld:2007} Auld, T., Bridges, M., 
Hobson, M.~P., \& Gull, S.~F.\ 2007, \mnras, 376, L11 

\bibitem[Azzalini \& Capitanio(2003)]{azzalini:2003} Azzalini, A. \&
  Capitanio, A. 2003, J.Roy.Statist.Soc, series B, vol.65, pp.367-389

\bibitem[Bennett et al.(2003)]{bennett:2003} Bennett, C.~L., et al.\ 
2003, \apjs, 148, 1 

\bibitem[Bond et al.(2000)]{bond:2000} Bond, J.~R., Jaffe, A.~H., 
\& Knox, L.\ 2000, \apj, 533, 19 

\bibitem[Chu et al.(2005)]{chu:2005} Chu, M., Eriksen, H.~K., Knox,
  L., G{\'o}rski, K.~M., Jewell, J.~B., Larson, D.~L., O'Dwyer, I.~J.,
  \& Wandelt, B.~D.\ 2005, \prd, 71, 103002

\bibitem[Eriksen et al.(2004)]{eriksen:2004} 
Eriksen, H.~K., et al.\ 2004, \apjs, 155, 227

\bibitem[Eriksen et al.(2007a)]{eriksen:2007a} Eriksen, H.~K., et al.\
2007a, \apj, 656, 641

\bibitem[Eriksen et al.(2007b)]{eriksen:2007b} Eriksen, H.~K., Huey, 
G., Banday, A.~J., G{\'o}rski, K.~M., Jewell, J.~B., O'Dwyer, I.~J., \& 
Wandelt, B.~D.\ 2007b, \apjl, 665, L1 

\bibitem[Eriksen et al.(2008a)]{eriksen:2008a} Eriksen, H.~K., Jewell, 
J.~B., Dickinson, C., Banday, A.~J., G{\'o}rski, K.~M., 
\& Lawrence, C.~R.\ 2008a, \apj, 676, 10 

\bibitem[Eriksen et al.(2008b)]{eriksen:2008b} Eriksen, H.~K., 
Dickinson, C., Jewell, J.~B., Banday, A.~J., G{\'o}rski, K.~M., 
\& Lawrence, C.~R.\ 2008b, \apjl, 672, L87 

\bibitem[Eriksen \& Wehus(2008)]{wehus:2008} Eriksen, H.~K. \&
  Wehus, I.~K.\ 2008, \apjs, in press, [arxiv:0806.3074]

\bibitem[Fendt \& Wandelt(2007)]{fendt:2007} Fendt, W.~A., \& Wandelt, B.~D.\ 2007, \apj, 654, 2 


\bibitem[Gelfand \& Smith(1990)]{gelfand:1990}
Gelfand, A. E., \& Smith, A. F. M. 1990, J. Am. Stat. Asso., 85, 398

\bibitem[Gelman \& Rubin(1992)]{gelman:1992}
Gelman, A., \& Rubin, D. 1992, Stat. Sci., 7, 457

\bibitem[Gold et al.(2008)]{gold:2008} Gold, B., et al.\ 2008, 
ApJS, [arXiv:0803.0715]

\bibitem[G{\'o}rski et al.(2005)]{gorski:2005} 
  G{\' o}rski, K.~M., Hivon, E., Banday, A.~J., Wandelt, B.~D.,
  Hansen, F.\,K., Reinecke, M., \& Bartelmann, M. 2005, \apj, 622, 759

\bibitem[Green \& Silverman(1994)]{green:1994} Green, P. J.,
  \& Silverman, B. W.\ 1994, Non-Parametric Regression and Generalized
  Linear Models, Chapman and Hall, 1994   

\bibitem[Groeneboom \& Eriksen(2008)]{groeneboom:2008} 
Groeneboom, N.~E., \& Eriksen, H.~K.\ 2008, ApJ, in press, [arXiv:0807.2242]

\bibitem[Hamimeche \& Lewis(2008)]{hamimeche:2008} Hamimeche, S., \&
  Lewis, A.\ 2008, \prd, 77, 103013  

\bibitem[Hinshaw et al.(2003)]{hinshaw:2003} Hinshaw, G., et al.\ 
2003, \apjs, 148, 63 

\bibitem[Hinshaw et al.(2007)]{hinshaw:2007} Hinshaw, G., et al.\ 
2007, \apjs, 170, 288 

\bibitem[Hinshaw et al.(2008)]{hinshaw:2008} Hinshaw, G., et al.\ 
2008, ApJS, in press, [arXiv:0803.0732]

\bibitem[Hivon et al.(2002)]{hivon:2002} 
Hivon, E., G{\' o}rski, K.~M., Netterfield, C.~B., Crill,
  B.~P.,Prunet, S., \& Hansen, F.\ 2002, \apj, 567, 2

\bibitem[Jewell et al.(2004)]{jewell:2004} 
  Jewell, J., Levin, S., \& Anderson, C.  H.  2004, \apj, 609, 1

\bibitem[Jewell et al.(2008)]{jewell:2008} 
  Jewell, J. B., et al. 2008, submitted, \apj, [astro-ph/0807.0624]

\bibitem[Komatsu et al.(2008)]{komatsu:2008} 
  Komatsu, E., et al.\ 2008, ApJS, in press, [arXiv:0803.0547]

\bibitem[Larson et al.(2007)]{larson:2007} Larson, D.~L., Eriksen,
H.~K., Wandelt, B.~D., G{\'o}rski, K.~M., Huey, G., Jewell, J.~B., \&
O'Dwyer, I.~J.\ 2007, \apj, 656, 653

\bibitem[Lewis \& Bridle(2002)]{lewis:2002} 
  Lewis, A., \& Bridle, S.\ 2002, \prd, 66, 103511

\bibitem[Nolta et al.(2008)]{nolta:2008} Nolta, M.~R., et al.\ 
2008, ApJS, in press, [arXiv:0803.0593]


\bibitem[O'Dwyer et al.(2004)]{odwyer:2004} O'Dwyer, I.~J., et al.\
2004, \apjl, 617, L99


\bibitem[Verde et al.(2003)]{verde:2003} Verde, L., et al.\ 2003, 
\apjs, 148, 195 

\bibitem[Wandelt et al.(2004)]{wandelt:2004} 
  Wandelt, B.~D., Larson, D.~L., \& Lakshminarayanan, A.\ 2004, \prd,
  70, 083511

\end{thebibliography}
\end{document}